\documentstyle[fleqn,prl,aps,preprint]{revtex}
\newcommand{\bfg}[1]{\mbox{\boldmath $#1$}}
\begin{document}
\title{
Non-Universal Fractional Quantum Hall States
in a \\ Quantum Wire
}
\author{Shinya {\sc Tokizaki} and Yoshio {\sc Kuramoto}}
\address{Department of Physics, Tohoku University, Sendai 980-77}
\maketitle

\begin{abstract}

The ground state as well as low-lying excitations
in a 2D electron system in strong magnetic fields
and a parabolic potential is investigated
by the variational Monte Calro method.
Trial wave functions analogous to the Laughlin state
are used with the power-law exponent as the variational parameter.
Finite size scaling of the excitation energy shows
that the correlation function at long distance is characterized by
a non-universal exponent in sharp contrast to
the standard Laughlin state.
The Laughlin-type state becomes unstable depending on
strength of the confining potential.

\end{abstract}
\clearpage

The fractional quantum Hall (FQH) effect occurs
in two-dimensional electron systems
under strong magnetic field \cite{Tsui}.
The ground state is described as an incompressible
quantum liquid with an energy gap \cite{Laughlin}.
On the edge of the FQH state, however, gapless excitations exist.
Hence the edge excitations play an important role in
low-energy response of the FQH system.
In a large system, the edge excitations are described
as the chiral Tomonaga-Luttinger (TL) liquid.
It has been discussed that the long-distance behavior
in the edge state is universal in the sense that
the exponent is determined by the filling alone
and is independent of other details of the system \cite{Wen}.

In a mesoscopic system where the length scale can be made
comparable to the radius of the cyclotron motion,
the FQH state should be affected by the finite dimension of the system.
In this paper we study the FQH state in a quantum wire
which can be fabricated as a conducting channel
in the semiconductor heterojunction.
If the channel is wide enough,
the chiral nature of the edge excitation should remain.
In a narrow channel, on the contrary,
two edges are not independent objects.
Then the universality of the asymptotic behavior might be lost.
The unique property of the quantum wire is that one can control
the strength of interaction between the edges,
and hence the chiral nature.
This  property is not shared with more popular
geometries like a disk or a semi-infinite plane.

The quantum wire in a strong magnetic field has been studied
theoretically by Yoshioka with use of the exact diagonalization method
\cite{Yoshioka}.
He has shown that a FQH state is realized under suitable condition
with respect to the strength of a confinement potential
and the Coulomb repulsion.
The exact diagonalization can deal with only small systems
containing several electrons.
As an alternative we use in this paper
the variational Monte Calro (VMC) method \cite{Mcmillan,Ceperley}
to investigate much larger systems with $O(10^2)$ electrons.
We aim at deriving explicitly the non-universal property of
the FQH state in the quantum wire, and show how the property depends on
the size of the system.
For this purpose we investigate not only the ground state
but low-lying excitations for the Laughlin-type wave
function \cite{Laughlin}.

Consider a two-dimensional electron system
under a strong magnetic field and a parabolic potential
in the $y$-direction.
There is no potential in the $x$-direction for
which we impose the periodic boundary condition with length $L_x$.
The magnetic field $B$ is applied antiparallel to the $z$-direction.
Then the one-electron Hamiltonian in the Landau gauge
is written as
\begin{equation}
 {\cal H_{\rm 0}}
   = \frac{1}{2 m_{\rm e}}
     \left[ (p_x+\frac{e}{c}By)^2 + p_y ^2 \right]
     + \frac{1}{2} m_{\rm e} \omega _0 ^2 y^2 \ ,
\end{equation}
where $m_{\rm e}$ is the electron mass
and $\omega_{\rm 0}$ is the strength of confinement potential.
The Hamiltonian is rewritten for later convenience as
\begin{equation} \label{eq:Ham_0}
   {\cal H_{\rm 0}} = - \frac{1}{2} \hbar \Omega
     \left [ \frac{1}{\mu} (\lambda \partial_x + i \frac{y}{\lambda} )^2
          + \mu \lambda^2 \partial_y^2  \right]
        - \frac{\hbar^2}{2 m_{\rm e}\lambda^2}
          \frac{\omega_0^2}{\Omega^2} \lambda^2 \partial^2_x \ .
\end{equation}
We have introduced
$\mu = \ell/\lambda = \sqrt{1+ (\omega_{\rm 0}/\omega_{\rm c})^2}$ and
$\Omega=\sqrt{\omega_{\rm c}^2 + \omega_{\rm 0}^2}$,
where $\ell = \sqrt{\hbar c/eB}$ and $\omega_c=eB/m_{\rm e} c$.
$\Omega$ and $\lambda$ give the effective cyclotron frequency
and the Larmor radius in the presence of the parabolic potential.
The eigenstate $\phi_{n,k}$ and the eigenvalue $E_{n,k}$
of this Hamiltonian are obtained as
\begin{equation}
\phi_{n,k}(\bfg{r}) =
\exp \left[ i k x-\frac{1}{2 \mu}
(\frac{y}{\lambda} + \lambda k)^2 \right]
{\rm H}_n(\frac{y+\lambda ^2 k}{\lambda\sqrt{\mu}}) \ ,
\end{equation}
\begin{equation}
E_{n,k}=\hbar \Omega ( n + \frac{1}{2}) +
   \frac{\hbar^2}{2 m_{\rm e}} \frac{\omega_0^2}{\Omega^2} k^2 \ .
\end{equation}
Here $k=2 \pi m /L_x$
($m$ is integer) is the momentum along the $x$-direction,
$n$ is the Landau level index and ${\rm H}_n$ is the Hermite polynomial.

The interaction between the electrons should also satisfy
the periodic boundary condition. We introduce the following function:
$$
V_{\rm c}(\bfg{r}) \equiv
\lambda
\left[ (\frac{L_{\rm x}}{\pi} \sin
 \frac{\pi}{L_{\rm x}}x)^2  + y^2\right] ^{-1/2}  \ ,
$$
which is periodic along the $x$-direction,
and reduces to the usual Coulomb form $\lambda/|\bfg {r}|$
in the limit of $x/L_x \to 0$.
Then the interaction Hamiltonian ${\cal H_{\rm int}}$ is given by
\begin{equation}
 {\cal H_{\rm int}} = \frac{e^2}{\varepsilon \lambda}
           \sum_{i<j} V_{\rm c}(\bfg{r}_{ij}), \hspace{1cm}
           \bfg{r}_{ij}=\bfg{r}_i - \bfg{r}_j \ ,
\end{equation}
where $\varepsilon$ is the dielectric constant.

{}From now on, we assume that the magnetic field is so strong
that it is sufficient to consider only the lowest Landau level ($n$=0),
with complete spin polarization.
The Laughlin-type wave function is constructed as follows:
\begin{equation}
\Psi_p (z_1,\ldots,z_N) =
\prod_{i<j} \left[ \sin \frac{\pi}{L_x}(z_i-z_j)
    \right]^p \prod_j \exp\left(-\frac{y_j^2}{2\mu \lambda^2} \right),
\label{eq:Laughlin}
\end{equation}
where is $z=x+iy/\mu$ and $p$ is an odd integer.
$\Psi_{p=1}$ is the Slater determinant of the non-interacting system,
and is the eigenstate of the $N$-body sum of eq.(\ref{eq:Ham_0}).
$\Psi_p$ with $p \neq 1$ is not an eigen function of
the non-interacting system in the presence of the parabolic potential.
However, the state belongs to the lowest Landau level.
Thus as long as $\omega_0/\omega_c \ll 1$,
$\Psi_p$ is expected to be a good trial function to account for
the repulsive correlation.
We take the number $N$ of electrons odd to avoid degeneracy
in the ground state.
Then the maximum momentum of the one-body state is
$\pi(N-1)p/L_x \sim \pi p\rho$
where $\rho=N/L_x$ is the line density.
Since the wave function with momentum $k$ is localized
around $-\lambda^2 k$ in the $y$-direction,
the width of the channel is about $p$ times of that with $p=1$.
Thus the Laughlin-type state corresponds to
the filling $\nu=1/p$ of the lowest Landau level.

The total energy of the trial state $\Psi_p$ is written as
\begin{equation}\label{eq:total}
E_{\rm 0} \sum_j
 \frac{\langle \Psi_p | -\lambda^2 \partial^2_{x_j} | \Psi_p \rangle }
       {\langle \Psi_p | \Psi_p \rangle}
      + U \sum_{i<j}
  \displaystyle{
   \frac{\langle \Psi_p | V_{\rm c} (\bfg{r}_{ij}) | \Psi_p \rangle}
        {\langle \Psi_p | \Psi_p \rangle} },
\end{equation}
where $E_{\rm 0}=(\hbar^2/2 m_{\rm e} \lambda^2)
(\omega_0^2/\Omega^2)=\hbar \omega_0^2/(2\omega_{\rm c})$
and $U=e^2/\varepsilon \lambda$ are constants characterizing
the magnitude of the confinement potential
and the Coulomb repulsion, respectively.
The term with $E_{\rm 0}$ in eq.(\ref{eq:total}) corresponds to
the final term in eq.(\ref{eq:Ham_0}).
We neglect the zero point oscillation
energy $\hbar \Omega N/2$
because it only shifts the origin of energy.
We calculate numerically these expectation values by the VMC method
taking $E_{\rm 0}/U \ (\ll 1)$
as the parameter to control the confinement.
For simplicity we put $\mu =1$,
which is justified in the case of
$\hbar \omega_{\rm 0} \ll e^2/(\varepsilon \ell)
\ll \hbar \omega_{\rm c}$.

We take the odd integer $p$ as the variational parameter
and optimize it for various values of $E_{\rm 0}/U$
and the line density $\rho$.
The results for $N$=45 is shown in Fig.\ref{fig:phase}.
The unit of length is taken to be $\lambda = \ell$ here and
from now on. In varying $p$ we keep $N$ constant,
so $\rho$ also represents the ratio of width to length of the system.
Hence the system with larger $\rho$ favors smaller $p$ as in
Fig.\ref{fig:phase}
because the effect of the external potential becomes larger.
Our result supports qualitatively the phase diagram obtained by Yoshioka.
In our case, however, more complex FQH states such as
$\nu = 2/3$ or $2/5$ have been neglected.
The presence of such states is indeed signaled as
an instability of the simplest Laughlin-type state
as will be discussed later.
Our primary concern in this paper, however,
is not the phase diagram itself.

We turn to excitation spectra from the Laughlin-type ground state,
and describe the results in terms of the TL liquid theory.
Since the one-body state in our model is characterized by
a single momentum along the $x$-direction,
the system can be regarded as quasi-1D.
The effective 1D field operator in the system is defined by
\begin{equation}
\Psi_1 (x) =
\frac{1}{\sqrt {2\pi} \lambda }\int_{-\infty}^{\infty}{\rm d}y
\Psi(\bfg r)\exp \left[-\frac{ix}{\lambda^2}(y-\frac{ix}{2})\right],
\end{equation}
where $\Psi(\bfg r)$ is the field operator in the lowest Landau level.
The integral corresponds to the average along
the $y$-direction with account of the cyclotron motion.
The Fourier transform of $\Psi_1(x)$ gives the annihilation
operator of electron with momentum $k$.
The low-energy excitations in a spinless 1D Fermion system
are classified into three kinds \cite{Haldane}:
(i) the particle-hole excitation giving rise to density fluctuation;
(ii) the number excitation associated with
the change of the chemical potential;
(iii) the current excitation with momentum twice the Fermi momentum.
Each excitation is gapless and is characterized
by the velocities $v_{\rm s}$, $v_N$ and $v_J$, respectively.
In terms of these velocities, the elementary excitation energies
are given by
(i) $2\pi v_{\rm s}m/L_x$ with $m$ positive integer,
(ii) $\pi (\Delta N)^2 v_N /2L_x$, and
(iii) $\pi J^2 v_J /2L_x$.
Here $J=N_{\rm R}-N_{\rm L}$ and $\Delta N=N'-N$ where
$N_{\rm R}$ ($N_{\rm L}$) is the number of right-mover (left-mover)
giving $N=N_{\rm R}+N_{\rm L}$, and where
 $N'$ is the particle number in the excited state.
If one can calculate the excitation energies,
these velocities are derived by analysis of their size dependence.

We shall derive the exponent which characterizes the asymptotic behavior
of the density matrix as
$
\langle \Psi_1^\dagger (x)\Psi_1(0)\rangle
\sim |x|^{-\theta-1}\cos(k_{\rm F} x)
$
where $k_{\rm F}$ is the Fermi momentum and
$ \theta = (K +1/K-2)/2 $.
It is known that
the exponent $K$ is derived from
$K=\sqrt{v_J/v_N}=v_J/v_{\rm s}$ \cite{Haldane}.
Of these velocities, $v_N$ involves the change of $N$ and
keeping the charge neutrality is a subtle procedure.
Therefore we use $v_J$ and $v_s$ to derive $K$.

Let us consider the following wave functions:
\begin{equation}
\Psi^{(\rm ph)}_p(z_1,\ldots,z_N) =
\sum_j \exp \left( i \frac{2 \pi}{L_x} z_j \right)
\Psi_p(z_1,\ldots,z_N)  \ ,
\end{equation}
\begin{equation}
\Psi^J_p(z_1,\ldots,z_N) =
\prod_j \exp \left( i \frac{2 \pi}{L_x} z_j \right)
\Psi_p(z_1,\ldots,z_N) \ .
\end{equation}
$\Psi_p^{(\rm ph)}$ describes the density excitation
with momentum $2\pi /L_x$ and gives the lowest excitation
of the type (i), while
$\Psi_p^J$ gives the lowest excitation of the type (iii) with $J$=2.
Both $\Psi_p^{(\rm ph)}$ and $\Psi_p^J$ are
orthogonal to $\Psi_p$,
namely $\langle \Psi_p^{(\rm ph)}|\Psi_p \rangle=0$ and
$\langle \Psi_p^J|\Psi_p \rangle=0$.
It is convenient to divide the excitation energy of a particle-hole pair from
the Laughlin-type state
eq.(\ref{eq:Laughlin}) into two parts:
$\Delta E^{(\rm ph)} = \Delta E_{\rm edge}^{(\rm ph)}
+ \Delta E_{\rm int}^{(\rm ph)}$.  Here $\Delta E_{\rm edge}^{(\rm ph)}$ comes
from the Hamiltonian part with $E_{\rm 0}$ and  $\Delta E_{\rm int}^{(\rm ph)}$
from that with $U$.
Fig.\ref{fig:fsc-ph} shows the VMC results in the case of $\rho=3.0$.
The size dependence is consistent with the gapless linear mode
expected from the TL theory.
The negative slope for the interaction part
$\Delta E_{\rm int}^{(\rm ph)}$ means
that the wider quantum wire reduces the Coulomb repulsion.
We can obtain numerically the velocity $v_{\rm s}$ from the slopes of
the lines in Fig.\ref{fig:fsc-ph}.
The results are summarized in Table \ref{tbl:velocity}.
On the other hand, $v_J$ can be calculated analytically
as $v_J=2\pi\rho E_{\rm 0}$ which is in fact independent of $p$ and $U$.

{}From the results of $v_s$ and $v_J$,
we can obtain the exponent $K$
using the relationship $K=v_J/v_{\rm s}$.
Fig.\ref{fig:K} shows the dependence of $K$ on
$E_{\rm 0}/U$ in the case of $\rho=3.0$ and $p=3$.
There appear three regions in Fig.\ref{fig:K}:
(I) $0.0013<E_{\rm 0}/U<0.0031$ ($K<0$),
(I{\kern-0.15em}I) $0.0031<E_{\rm 0}/U<0.0047$ ($K>1$) and
(I{\kern-0.15em}I{\kern-0.15em}I)
$0.0047<E_{\rm 0}/U<0.0085$, ($0<K<1$).
The excitation energy is negative in the case of $K<0$. This means
that $\nu=1/3$ Laughlin-type state is unstable in that region (I).
Unfortunately it is impossible to obtain the true ground state
in our method.
In the region of (I{\kern-0.15em}I) and
(I{\kern-0.15em}I{\kern-0.15em}I),
the excitation energy is positive,
and the system behaves as a TL liquid.
In the region (I{\kern-0.15em}I) the pairing fluctuation is dominant
with $K>1$.
One might interpret this as follows:
The Coulomb repulsion is overcompensated in making the Laughlin-type state, and
the remaining correlation becomes attractive.
It is however more likely that the real ground state
in that region is not the $\nu =1/3$ state.
The most interesting region is (I{\kern-0.15em}I{\kern-0.15em}I)
where $K$ approaches $1/3$ from above as $U$ gets smaller.
This behavior is in contrast with the universal exponent $K=1/p$
in a large FQH droplet with $\nu=1/p$ \cite{Wen,Stone&Fisher}.
The deviation is interpreted in terms of interaction between the edges.
As the interaction becomes insignificant with the decrease of $U$,
the exponent approaches the universal value.
If $U$ becomes too small, however, the 1/3 state becomes unstable.
We have confirmed that the $\nu=1/5$ state has a similar character
with respect to $K$ as the $\nu=1/3$ state.
In the limit of large width,
the density matrix should decay as
$|x|^{-1/\nu}$ according to the chiral Luttinger liquid theory \cite{Wen}.
Here the distance $|x|$ should be smaller than the width of the system,
and the field operator consists of right- and left-going components
representing both edges.

In summary we have shown that the $\nu=1/p$
Laughlin-type state in a quantum wire has the edge excitation
which has a non-universal exponent dependent on
details of the system.
We have demonstrated that competition
between the Coulomb repulsion and the harmonic confinement
determines the value of $p$ in the ground state.
The next problem to be investigated is how the non-universality appears in the
response of the system such as the Hall coefficient and in the temperature
dependence of the conductance.

The authors would like to thank H. Yokoyama and Y. Kato
for helpful discussion.
One of the authors (S.T) gratefully acknowledges
the guidance of H. Ebisawa and K. Yonemitsu.

\begin{figure}[r]

\caption{Phase diagram of the system with $N=45$.
This result is an average over 5$\times 10^4$ samples.}
\label{fig:phase}
\vspace{3cm}
\end{figure}

\begin{figure}[r]
\caption{(a) $\Delta E_{\rm edge}^{\rm (ph)}/L_x E_{\rm 0}$ and
(b) $\Delta E_{\rm int}^{(\rm ph)}/L_x U$ as a function of $1/L_x^2$
for particle--hole excitation from $\nu=1/p$ Laughlin state.
The line density is fixed to be
$\rho=3.0$, so that $N$ is varied up to 75.
The result of $\nu=1$, 1/3 and 1/5 are shown with open circle,
square and triangle respectively.
These excitation energies are obtained by averaging over
1 $\times 10^5$ samples. }
\label{fig:fsc-ph}
\vspace{3cm}
\end{figure}

\begin{figure}[r]
\caption{The exponent $K$ of TL-liquid by as a function of
$E_{\rm 0}/U$ for $\rho=3.0$ and $\nu=1/3$.
The range of $E_{\rm 0}/U$ is taken so that
the $\nu=1/3$ Laughlin-type state is the
ground state for $N= 45$.
The regions
(I), (I{\kern-0.15em}I) and (I{\kern-0.15em}I{\kern-0.15em}I)
have $K<0$, $K>1$ and $0<K<1$, respectively.
$K=1/3$ is the value expected for the chiral TL liquid.}
\label{fig:K}
\vspace{3cm}
\end{figure}

\begin{table}
\caption{Velocity of the particle-hole excitation from $\nu=1/p$
Laughlin state.}

\label{tbl:velocity}
   \begin{tabular} {cc}
   $\nu$     & $v_{\rm s} $ \\
   \hline
    \ \ \ 1 \ \ \   &
    $1.844 \times 10^1    E_{\rm 0}
    -5.134 \times 10^{-1} U         $\\
    \ \ \ 1/3 \ \ \ &
    $5.574 \times 10^1    E_{\rm 0}
    -1.727 \times 10^{-1} U         $\\
    \ \ \ 1/5 \ \ \ &
    $9.163 \times 10^1    E_{\rm 0}
    -9.996 \times 10^{-2} U         $
   \end{tabular}
\end{table}

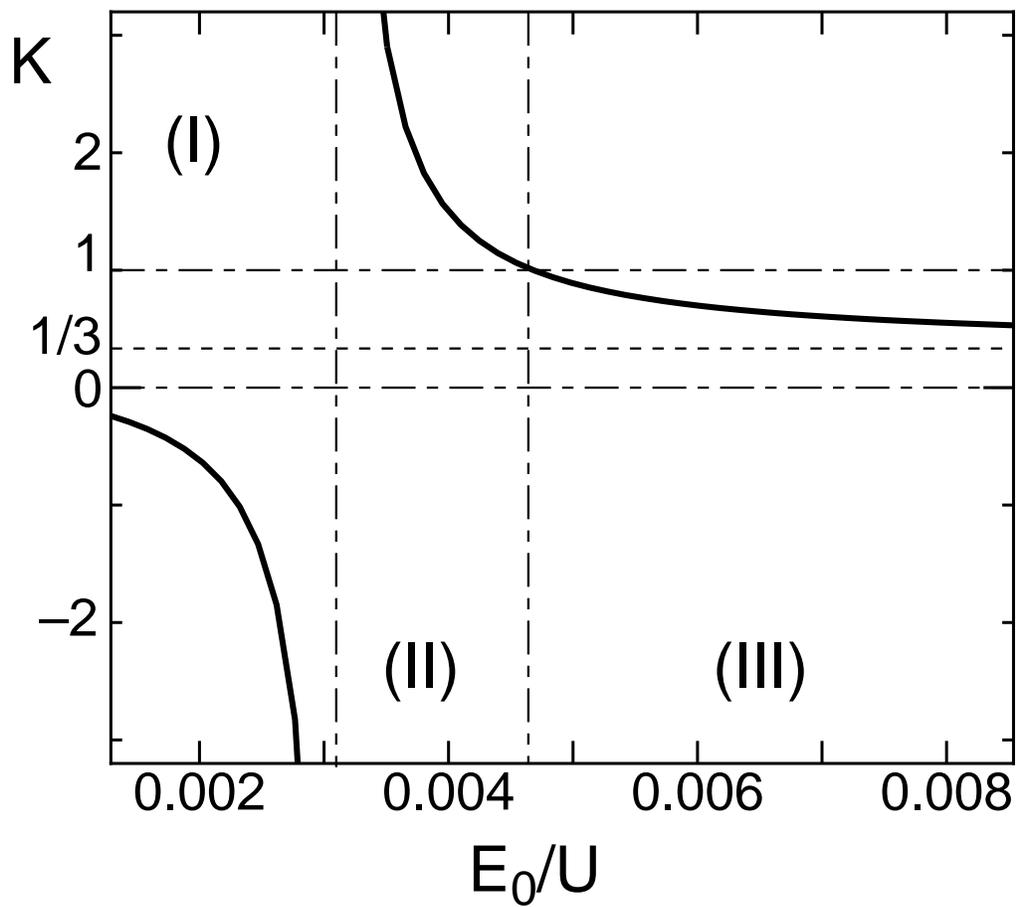
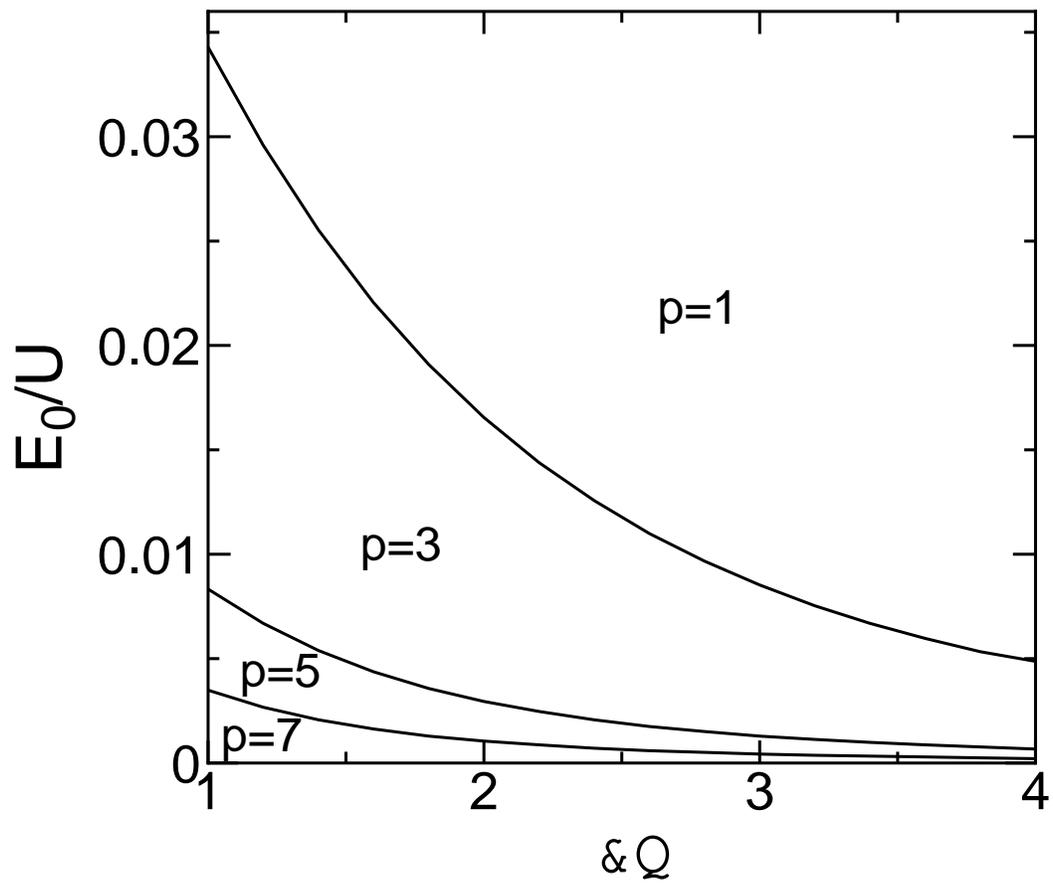
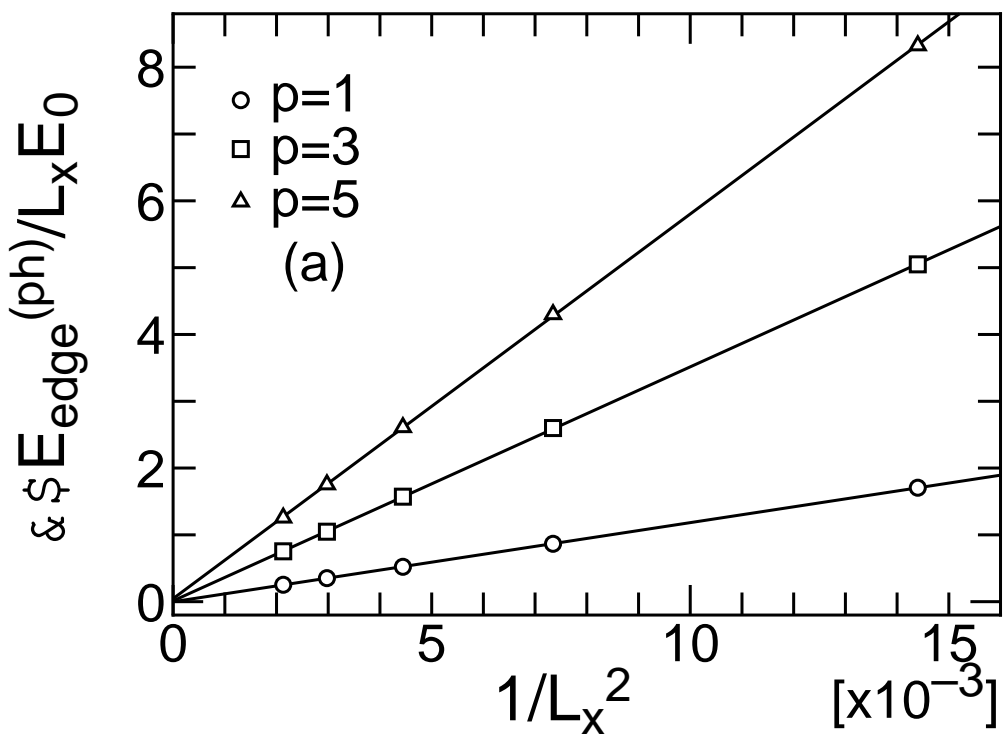
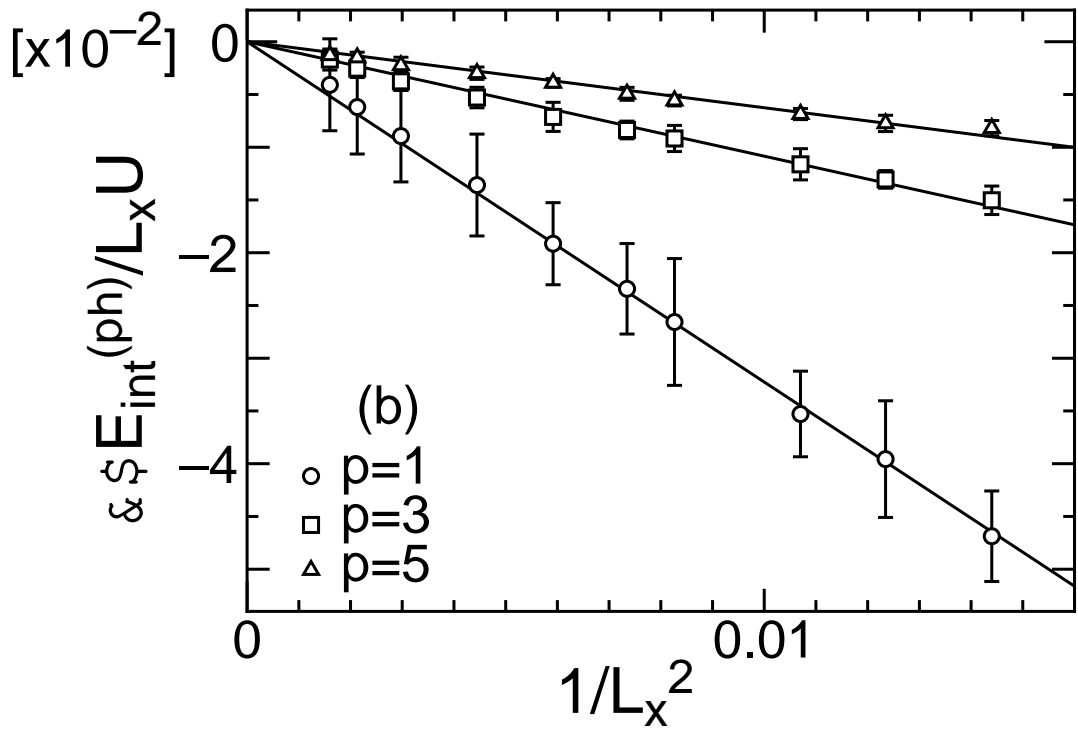
\end{document}